\documentclass[floatfix,aps,pre,showpacs,showkeys,groupedaddress]{revtex4}
\usepackage{graphicx}
\usepackage{amsmath}
\begin{document}

\preprint{CAS-ITP-2006-212-04}

\title{Boltzmann distribution of free energies in a finite-connectivity spin-glass system and the cavity approach\footnote{Citation information: Frontiers of Physics in China {\bf 2}: 238-250 (2007).} }

\author{Haijun Zhou}

\affiliation{Institute of Theoretical Physics, the Chinese Academy of 
  Sciences, Beijing 100080, China}

\date{April 06, 2007}

\begin{abstract}
At sufficiently low temperatures, the configurational phase space of a large spin-glass system breaks into many separated domains, each of which is referred to as a macroscopic state. The system is able to visit all spin configurations of the same macroscopic state, while it can not spontaneously jump between two different macroscopic states. Ergodicity of the whole configurational phase space of the system, however, can be recovered if a temperature-annealing process is repeated an infinite number of times. In a heating-annealing cycle, the environmental temperature is first elevated to a high level and then decreased extremely slowly  until a final low temperature $T$ is reached. Different macroscopic states may be reached in different rounds of the annealing experiment; while the probability of finding the system in macroscopic state $\alpha$ decreases exponentially with the free energy $F_\alpha(T)$ of this state.  For finite-connectivity spin glass systems, we use this free energy Boltzmann distribution to formulate the cavity approach of M{\'{e}}zard and Parisi [Eur.~Phys.~J.~B {\bf 20}, 217 (2001)] in a slightly different form. For the $\pm J$ spin-glass model on a random regular graph of degree $K=6$, the predictions of the present work agree with earlier simulational and theoretical results.
\end{abstract}


\maketitle

\section{Introduction}
\label{sec:introduction}
Spin-glasses are simple models for disordered systems. They can be defined very easily in mathematical terms; on the hand, the properties of such simple models usually are quite rich. Statistical physics of spin-glasses has been studied for more than thirty years since the concept of spin-glasses was first presented by Edwards and Anderson \cite{Edwards-Anderson-1975} in 1975, but there are still many unsolved and heavily debated issues. In the last decades there have been a lot of theoretical investigations concerning models defined on a  finite-connectivity random graph.  These later models are more realistic than conventional spin-glass models (e.g., the Sherrington-Kirkpatrick model \cite{Sherrington-Kirkpatrick-1975}) on a complete graph, in the sense that each spin interacts only with a finite number of other spins.  As direct analytical studies of spin-glass models on three-dimensional (regular) lattices are still beyond reach, people hope that a deep understanding of 
(mean-field) models on finite-connectivity random graphs will shed much light on the properties of 3D  systems.

For a spin-glass system of very large size, it is generally believed that ergodicity is broken when the environmental temperature $T$ becomes lower than a certain value $T_{\rm sg}$, the spin-glass transition temperature. In this low-temperature spin-glass phase, the time average of a given physical quantity no longer equals to the ensemble average. The whole configurational phase space of the system breaks into many separated domains. Ergodicity is still preserved within each such configurational domains. If the system initially is in a spin configuration that belongs to domain $\alpha$ of the configurational space, it will eventually visit all other configurations in this domain as time elapses. On the other hand, due to the existence of very high free energy barriers, the system is unable to transit spontaneously from one  configurational space domain to another different domain. In this ergodicity-broken situation, each such configurational phase space domain is regarded as a {\em macroscopic state} or {\em thermodynamic state}.  The free energy $F_\alpha$ of a macroscopic state $\alpha$ is related to the microscopic configurations in $\alpha$ through the following fundamental formula of statistical mechanics,
\begin{equation}
  \label{eq:Falpha}
   F_\alpha(N, \beta) =  -T \ln Z_\alpha(N, \beta) \ ,
\end{equation}
where $N$ is the total number of spins in the system; $\beta = 1/T$ is the inverse 
temperature (we set Boltzmann's constant to unity throughout this paper); 
and $Z_\alpha(N, \beta)$ is the partition function for the macroscopic state $\alpha$ as defined by
\begin{equation}
\label{eq:PartitionFunctionAlpha}
Z_\alpha(N,\beta)= \sum\limits_{\vec{\sigma } \in \alpha} \exp\bigl[-\beta H(\vec{\sigma })\bigr] \ .
\end{equation}
In Eq.~(\ref{eq:PartitionFunctionAlpha}), $\vec{\sigma}$ denotes a microscopic spin configuration, and $H(\vec{\sigma})$ is the total energy of this configuration. The summation in Eq.~(\ref{eq:PartitionFunctionAlpha}) is over all those microscopic configurations belonging to macroscopic state $\alpha$. 

For a given spin-glass system, although we can formally write down the expressions of the partition function and free energy for a  macroscopic state $\alpha$, the principal difficulty of spin-glass statistical physics is that we do not know {\em a priori} how the configurational space of the system is organized and which are the constituent spin configurations of each macroscopic state. To overcome this difficulty, one possibility is to first assume certain structural organization of the system's configurational space and then try to derive a self-consistent theory. For the Sherrington-Kirkpatrick model, the full-step replica-symmetry-broken (FRSB) theory of Parisi with an ultrametric organization of macroscopic states \cite{Mezard-etal-1987} has met with great success. For finite-connectivity mean-field spin-glasses, a cavity approach was also developed by M{\'{e}}zard and Parisi \cite{Mezard-Parisi-2001}. This cavity approach combined the Bethe-Peierls approximation \cite{Bethe-1935,Peierls-1936,Peierls-1936a} for a ferromagnetic Ising model with the physical picture that there is a proliferation of macroscopic states in a spin-glass system. This approach, which has been shown \cite{Mezard-Parisi-2001} to be equivalent to the first-step replica-symmetry-broken (1RSB) replica theory, can give very good predictions concerning the low-temperature free energy density of a system on a random graph. Later on, this cavity approach was extended by M{\'{e}}zard and Parisi \cite{Mezard-Parisi-2003} to the limiting case of zero temperature. The zero-temperature cavity method was applied to some hard combinatorial optimization problems with good performances (see, e.g., Refs.~\cite{Mezard-etal-2002,Mezard-Zecchina-2002,Braunstein-etal-2005,Weigt-Zhou-2006}). These interdisciplinary applications in return also call for further understanding on the mean-field cavity approach and its possible extensions (see, e.g., Refs.~\cite{Zhou-2005a,Zhou-2005b,Montanari-Rizzo-2005,Parisi-Slanina-2006,Chertkov-Chernyak-2006a,Chertkov-Chernyak-2006b}).

For a spin-glass system at a low temperature $T$, the total number of macroscopic states with a given free energy $F$ is denoted as $\Omega_{\rm mac}(F)$.  Although it is natural to anticipate that the logarithm of $\Omega_{\rm mac}(F)$ should be an extensive quantity, the exact relationship between $\Omega_{\rm mac}$ and $F$ is unknown and is system-dependent. Inspired by Parisi's FRSB solution, in the cavity approach of M{\'{e}}zard and Parisi \cite{Mezard-Parisi-2001} one assumes that the total number ${\Omega}_{\rm mac}(F)$ of macroscopic states with free energy $F$ diverges exponentially with $F$ with respect to a reference free energy $F_r$ \cite{Mezard-etal-1987}, i.e.,
\begin{equation}
  \label{eq:MP01}
  {\Omega}_{\rm mac}(F) \propto \exp\Bigl( x \beta (F-F_r) \Bigr) \ ,
\end{equation}
where $0 < x < 1$ is a dimensionless constant to be determined self-consistently. The Parisi parameter $x$ stems from the replica theory of infinite-connectivity spin-glasses \cite{Mezard-etal-1987}, its physical meaning in the cavity framework is not transparent.  For some spin-glass systems with many-body interactions, even if the exponential form of Eq.~(\ref{eq:MP01}) is valid in certain range of free energy values, the parameter $x$ in this equation may exceed unity. The original cavity iterative equations of Ref.~\cite{Mezard-Parisi-2001} diverge for this situation of $x\geq 1$, where the statistical physical property of the spin-glass system is not determined by those macroscopic states with the globally minimal free energy density.  Since the cavity approach of M{\'{e}}zard and Parisi is a very useful theoretical tool in studying the low-temperature properties of many spin-glass systems, it might be helpful for us to interpret the approach from an another slightly different angle. Complementary interpretations of the 1RSB cavity theory will also facilitate its further development.

In this paper we demonstrate that, the 1RSB cavity formalism of M{\'{e}}zard and Parisi can be derived in an alternative way without using Eq.~(\ref{eq:MP01}).  This slightly revised cavity theory is based on the gedanken experiment of repeated temperature heating-annealing. The final macroscopic state reached at the end of an annealing process is anticipated to follow the Boltzmann distribution of free energies.  This theoretical approach is applied to the $\pm J$ spin-glass model on a random regular graph of degree $K=6$ to test its validity. The results of the present work are in close agreement with earlier simulational and numerical results. The mathematical format of the present cavity theory is the same both for non-zero temperatures and for zero temperature. If the macroscopic states of a spin-glass system further organize into clusters of macroscopic states, it can be easily extended to take into account this situation.

This paper is organized as follows. The next section introduces the $\pm J$  spin-glass model and the ensemble of random regular graphs of degree $K$. Section \ref{sec:mfep} describes the free energy Boltzmann distribution of macroscopic states. In Sec.~\ref{sec:cavity-field-distribution} the concepts of cavity field and cavity magnetization are re-introduced, and the distribution of a vertex's cavity magnetization among all the macroscopic states is calculated. Various thermodynamic quantities, including the grand free energy density of the
whole system and the free energy density of a macroscopic state, are calculated in Sec.~\ref{sec:densities} and Sec.~\ref{sec:single-instance} for an ensemble of systems and for a single system, respectively. The numerical results for the $\pm J$ spin-glass model are reported and analyzed in Sec.~\ref{sec:numerics}. We conclude the present work and discuss its possible extensions in Sec.~\ref{sec:discussion}. The appendix gives an explicit expression for the mean energy density.

\section{The  $\pm J$ spin-glass model on a random regular graph}
\label{sec:model}

In this paper we focus on just a single example, the $\pm J$ spin-glass model \cite{Viana-Bray-1985} on a random regular graph of degree $K$. This model was also studied in Ref.~\cite{Mezard-Parisi-2001}, lending us the opportunity to directly compare the results of both treatments.

Let us consider the graph ${\cal G}_K(N)$ which is obtained by randomly choosing with equal probability a graph from the set of all regular graphs of size $N$ and vertex-degree $K$. ${\cal G}_K(N)$ is a random regular graph of degree $K$. Each of the $N$ vertices of ${\cal G}_K(N)$ is connected to $K$ other vertices, but there is no structure in the connection pattern of ${\cal G}_K(N)$. For each edge $(i,j)$ of graph ${\cal G}_K(N)$, we assign a coupling constant $J_{i j}= + J$ or $J_{i j}= - J$ with equal probability. Once the coupling $J_{i j}$ for an edge is assigned, it no longer changes. Therefore, we have a {\em network} with random quenched connection pattern and random quenched coupling constants.  (In the remaining part of this paper, when we use the term $`$network', we mean the connection pattern of the graph plus the quenched couplings; when we use the term $`$graph', we mean only the connection pattern.) On top of such a network we define the following energy function
\begin{equation}
  \label{eq:energy01}
  H(\sigma_1, \sigma_2, \ldots, \sigma_N; {\cal G}) =
 - \sum\limits_{(i,j) \in {\cal G}} J_{i j} \sigma_i \sigma_j \ ,
\end{equation}
where $\sigma_i$ ($i=1,\ldots, N$) is the spin variable of vertex $i$, $\sigma_i$ can take two values, $\sigma_i=-1$ and $\sigma_i = + 1$.  Equation (\ref{eq:energy01}) is the $\pm J$ spin-glass model. The total number of microscopic configurations $\vec{\sigma}=\{ \sigma_1, \sigma_2, \ldots, \sigma_N \}$ in such a system is simply $2^{N}$.

The vertex degree of a random regular graph of degree $K$  is equal to that of a regular square lattice in $d=K/2$ dimensions. One hopes that some statistical physical properties of a spin-glass model on a random regular graph will also hold for the same model on a regular lattice. On the other hand, in a regular lattice, there exist many short loops, which make analytical calculations extremely difficult. In a random regular graph of very large size there is no such short loops.  The graph is locally tree-like; and the typical loop length in the graph scales as $\log_{K-1}(N)$. One can exploit this absence of short loops to construct a self-consistent mean-field theory.

For the benefit of later discussions, we also mention the concept of a random regular cavity graph ${\cal G}_K(N,m)$ \cite{Mezard-Parisi-2001}. In this graph of size $N$, there are $m$ ($m\ll N$) vertices of degree $K-1$ and $N-m$ vertices of degree $K$. A vertex of degree $K-1$ in graph ${\cal G}_K(N,m)$ is referred to as a {\em cavity vertex}. Like ${\cal G}_K(N)$, the connection pattern of ${\cal G}_K(N,m)$ is also completely random; and the quenched coupling constants for the edges of ${\cal G}_K(N,m)$ are also independently and  equally distributed over $\pm J$.  The spin-glass Hamiltonian Eq.~(\ref{eq:energy01}) can also be defined for this cavity network.

\section{The gedanken annealing experiment and the Boltzmann distribution of free energies}
\label{sec:mfep}

\begin{figure}[ht]
  \includegraphics[width=0.6\linewidth]{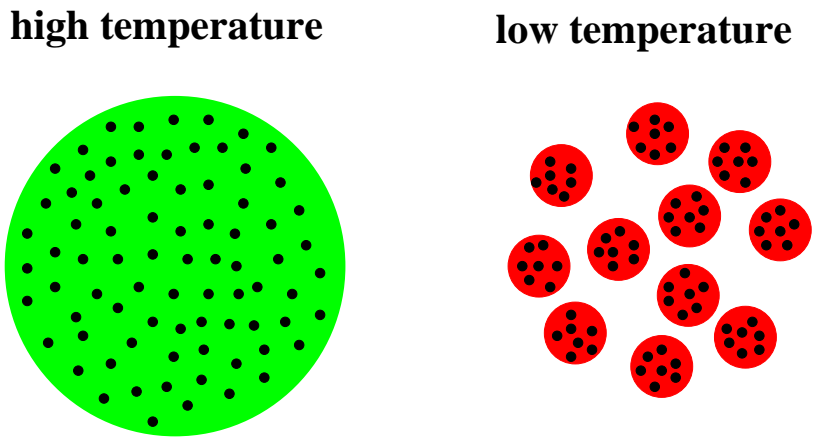}
  \caption{\label{fig:ErgodicityBreaking01}
A schematic picture of ergodicity breaking in a spin-glass system. (Left) When the temperature is  high,  all the microscopic configurations (denoted by dots) which contribute to the free energy of the system are in a single domain of the  configurational phase space. (Right) when the temperature is low enough, the microscopic configurations which contribute to the free energy of the system are contained in different domains in the configurational phase space. Each domain contains a number of microscopic configurations. Ergodicity is still preserved within each phase space domain, while the configuration of the system is unable to jump spontaneously from one domain to another different domain of the phase space.
}
\end{figure}

Consider the spin-glass system Eq.~(\ref{eq:energy01}). At high temperatures, the system is in the paramagnetic phase. Each vertex of the network does not have any spontaneous magnetization, its spin fluctuates over the positive and negative directions and stays in each orientation with equal probability. The configurational phase space of the system is ergodic (see left panel of Fig.~\ref{fig:ErgodicityBreaking01}).  When the temperature $T$ is decreased below a spin-glass transition temperature $T_{\rm sg}$, the system is in the spin-glass phase. In this phase, a vertex of the system might favor one spin orientation over the opposite orientation, but this orientation preference is vertex-dependent.  When the number $N$ of vertices is sufficiently large, the configurational phase space of this spin-glass system splits into many separated domains (see right panel of Fig.~\ref{fig:ErgodicityBreaking01}). Each domain of the phase space, which corresponds to a macroscopic state or thermodynamic state of the system, contains a set of microscopic spin configurations. The system is ergodic within each macroscopic state, while ergodicity is broken at the level of macroscopic states.   However, the system is able to transit from one macroscopic state to another different macroscopic state if the following temperature heating-annealing process is performed: The system is first heated to a high temperature beyond $T_{\rm sg}$ and  waited for a long time till it reaches equilibrium. The system now stays in a high-temperature ergodic phase. Afterwords,  it is cooled infinitely slowly till the final low temperature $T$ is reached. (A simulated-annealing idea was previously explored by Kirkpatrick and co-workers \cite{Kirkpatrick-etal-1983} to tackle hard optimization problems.)  Since in the high-temperature phase the system loses memory about its prior history, at the end of the annealing process it may reach any a macroscopic state.  The probability $P_\alpha$ of the system end up being in a particular macroscopic state $\alpha$, however, is in general different for different macroscopic states.  We argue in the following that, if an infinite number of the above-mentioned annealing experiment are performed, $P_\alpha$  should depend only on the free energy $F_\alpha$ of the macroscopic state $\alpha$.

The total partition function of the system is defined as
\begin{equation}
  \label{eq:PartitionTotal}
  {\cal Z}(N,\beta) =
\sum\limits_{\vec{\sigma}} \exp\bigl(-\beta H( \vec{\sigma}) \bigr) \ ,
\end{equation}
where $H(\vec{\sigma})$ is the total energy expression as given by Eq.~(\ref{eq:energy01}). When ergodicity is broken, the partition function ${\cal Z}(N,\beta)$ can be re-written as a sum over all the macroscopic states $\alpha$:
\begin{equation}
  {\cal Z}(N,\beta) = \sum\limits_{\alpha} \sum\limits_{\vec{\sigma} \in \alpha} 
    \exp\bigl( -  \beta H( \vec{\sigma}) \bigr)
  = \sum\limits_{\alpha} Z_\alpha(N,\beta) 
  = \sum\limits_{\alpha} \exp\biggl( - \beta  F_\alpha( N , \beta) \biggr) \ .
  \label{eq:PartitionDomain02} 
\end{equation}
In Eq.~(\ref{eq:PartitionDomain02}), $Z_\alpha(N, \beta)$ and $F_\alpha (N, \beta)$ are the partition function and free energy of the macroscopic state $\alpha$ as defined in Eq.~(\ref{eq:PartitionFunctionAlpha}) and Eq.~(\ref{eq:Falpha}), respectively. By way of repeated temperature annealing, the whole configurational space of a spin-glass system is explored (and ergodicity is recovered!). The total partition function Eq.~(\ref{eq:PartitionTotal}) can therefore be understood as containing all the information of the system as measured by an infinite number of temperature annealing experiments. From Eq.~(\ref{eq:PartitionDomain02}) we see that each macroscopic state $\alpha$ contributes a term  $e^{-\beta F_\alpha(N, \beta)}$ to the total partition function of the system.  Therefore we can anticipate that, at temperature $T$,  the probability of the system being in macroscopic state $\alpha$ at the end of an annealing experiment is  given by the following distribution
\begin{equation}
  \label{eq:BoltzmannDistribution}
  P_\alpha(\beta) =\frac{ e^{-\beta F_\alpha(N, \beta) }}{  \sum_\alpha 
  e^{-\beta F_\alpha (N, \beta)} } \ .
\end{equation}
This is a free energy Boltzmann distribution over all the macroscopic states $\alpha$.

At this point, we introduce an artificial inverse temperature $y$ at the level of 
macroscopic states. (Such an artificial inverse temperature $y$ was first introduced in the early work of Ref.~\cite{Mezard-Parisi-2003} for $T=0$ spin-glasses.) This is better explained by constructing the following artificial single-particle system:  the particle has a set of  $`$energy' levels, the $`$energy' of level $\alpha$ is equal to the free energy $F_\alpha(N, \beta)$ of the macroscopic state $\alpha$ of the actual spin-glass system; this one-particle system is in a heat bath with inverse temperature $y$. Then the partition function of the artificial system is
\begin{equation}
  \label{eq:NewPartitionFunction}
  {\cal Z}(N, \beta; y) = \sum\limits_{\alpha} \exp\biggl( - y F_\alpha(N, \beta)
   \biggr) \ .
\end{equation}
The probability of this artificial system being in energy level $\alpha$ is given by
\begin{equation}
  \label{eq:BoltzmannDistribution02}
 P_\alpha(\beta; y) = \frac{ e^{-y F_\alpha(N, \beta) } }
{ \sum_\alpha e^{-y F_\alpha(N, \beta)} } \ .
\end{equation}
When the inverse temperature $y$ of the artificial system is set to $y=\beta$, then the partition function ${\cal Z}(N, \beta; y)$ reduces to the total partition function ${\cal Z}(N, \beta)$ of the original spin-glass system.

The partition function ${\cal Z}(N, \beta; y)$ of Eq.~(\ref{eq:NewPartitionFunction})
can be written in another form as
\begin{equation}
  {\cal Z}(N, \beta; y) = \sum\limits_{\alpha} \exp\bigl(- y F_\alpha \bigr)
  = \sum\limits_{F} \Omega_{\rm mac}(F) e^{-y F} 
  = \sum\limits_{F} \exp\biggl[ N \bigl( \Sigma(f) - y f \bigr) \biggr] \ .
  \label{eq:Zyb03}
\end{equation}
In Eq.~(\ref{eq:Zyb03}), $f=F/N$ is the free energy density of a macroscopic state; the function $\Sigma(f)$ is called the complexity \cite{Mezard-Parisi-2001,Mezard-Parisi-2003}, which is related to the total number $\Omega_{\rm mac}(F)$ of macroscopic states by the following equation
\begin{equation}
  \label{eq:cp01}
  \Sigma(f) = \frac{1}{ N} \ln \Omega_{\rm mac}(N f) \ .
\end{equation}
The complexity $\Sigma(f)$, which should be non-negative, is a measure of the total number of macroscopic states with free energy density $f$.  When the spin-glass system is very large ($N \gg 1$), Eq.~(\ref{eq:Zyb03}) indicates that  the partition function ${\cal Z}(N, \beta; y)$  is contributed exclusively by those macroscopic states whose free energy density $f$ satisfying the equation
\begin{equation}
  \label{eq:y}
  \frac{ \partial \Sigma(f) }{ \partial f}  = y \ .
\end{equation}
Equation (\ref{eq:y}) gives an implicit relationship between the inverse temperature $y$ and the observed mean free energy density $f$ of the system.

For the benefit of later discussions, we define a grand free energy $G(N, \beta;f)$ by the following equation
\begin{equation}
 \label{eq:grand_free_energy}
G(N, \beta; y) \equiv -\frac{1}{y} \ln {\cal Z}(N, \beta; y)
= -\frac{1}{y} \ln\biggl[\sum\limits_{\alpha}\exp\bigl( - y F_\alpha \bigr) \biggr]\ .
\end{equation}
From Eq.~(\ref{eq:Zyb03}) we see that the grand free energy density $g(\beta;y)$ is
\begin{equation}
\label{eq:grand_free_energy_density_01}
g(\beta;y)\equiv \lim\limits_{N\to \infty} \frac{G(N, \beta; y)}{N} = \min\limits_{f} \bigl( f - \Sigma(f)/ y \bigr) = f(\beta;y)-\Sigma\bigl(f(\beta;y)\bigr)/y \ ,
\end{equation}
where $f(\beta;y)$ is the solution of Eq.~(\ref{eq:y}).

In the next three sections we will study the statistical physical properties of the $\pm J$ spin-glass system  using $\beta$ and $y$ as a pair of control parameters. The free energy at the microscopic level (within a macroscopic state) will be calculated with the inverse temperature $\beta$, while the grand free energy at the macroscopic state level will be calculated with the inverse temperature $y$. Although in the actual spin-glass system, the two inverse temperatures are identical ($y\equiv \beta$), they are decoupled in the following analytical theory. This decoupling gives us extra freedom in the theoretical development. Finally, to go from the artificial system to the original system, we will choose the largest value of $y$ in the interval of $ 0 \leq y \leq \beta$ which satisfies the requirement that the complexity $\Sigma$ is non-negative.

To analytically study the statistical physical property of a spin-glass model on 
a random graph, there are basically two different but equivalent approaches. The first one is the replica method \cite{Viana-Bray-1985,Kanter-Sompolinsky-1987,Monasson-1998} and the second one is the cavity method \cite{Mezard-Parisi-2001}. The present paper exploits the 
cavity approach. It corresponds to the 1RSB approximation of the replica method. Basically, one assumes that macroscopic states of the spin-glass system are distributed evenly in the whole configurational phase space, and there is no further organization of the macroscopic states or further structures within each macroscopic state.

\section{Integrating the Boltzmann distribution of free energies into the cavity approach}
\label{sec:cavity-field-distribution}

\subsection{Cavity fields}
\label{subsec:single01}

\begin{figure}[ht]
  \begin{center}
    \includegraphics[width=0.7\linewidth]{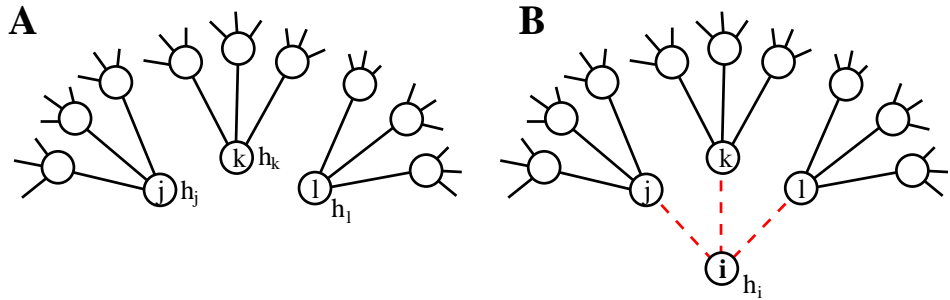}
  \end{center}
  \caption{\label{fig:cavity01m1}
Cavity vertices in a random regular cavity graph of vertex degree $K$ (here $K=4$). (A) The neighborhoods for three cavity vertices $j$, $k$ and $l$. These three vertices  have vertex degree $K-1$ ($=3$) while all the other $N-3$ vertices in the graph have vertex degree $K$. In a macroscopic state $\alpha$, the three cavity vertices feel the cavity fields $h_j$, $h_k$ and $h_l$, respectively. (B)  A new cavity graph with $N+1$ vertices is generated by adding a new vertex $i$ and connecting it to $K-1$ old cavity vertices. The cavity vertex $i$ of the new graph feels the cavity field $h_i$ in the corresponding macroscopic state $\alpha$ of the new cavity system.
}
\end{figure}

Consider the spin-glass system Eq.~(\ref{eq:energy01}) on a cavity network ${\cal G}_K(N,K-1)$ of $N$ vertices and $K-1$ cavity vertices (see Fig.~\ref{fig:cavity01m1}A). Let us suppose that the configurations of such a cavity system are in a macroscopic state $\alpha$.

At given inverse temperature $\beta$, the spin value $\sigma_j$ of a cavity vertex $j$ fluctuates over time around certain mean value $m_j$ which depends on the macroscopic state $\alpha$. Since $\sigma_{j}$ is  a binary variable, the marginal distribution $ \rho_j(\sigma_j)$ of $\sigma_j$ can be expressed in the following form %
\begin{equation}
  \label{eq:CavityField01}
\rho_j(\sigma_{j}) = \frac{ e^{ \beta h_{j} \sigma_{j}} }{ \cosh \beta h_{j} }
\ ;
\end{equation}
the magnetization of the cavity vertex $j$ is
\begin{equation}
  \label{eq:Magnetization01}
  m_{j} = \tanh \beta h_{j} \ .
\end{equation}
From Eq.~(\ref{eq:Magnetization01}) we know that $h_j$ is the magnetic field experienced by the cavity vertex $j$ due to the spin-spin interactions between vertex $j$ and its $K-1$ nearest-neighbors. We call $h_j$ the {\em cavity field} on vertex $j$ and $m_j$ the {\em cavity magnetization} of vertex $j$. We emphasize again that $h_j$ and $m_j$ depends on the macroscopic state $\alpha$.

Since the connection pattern in a cavity graph with $K-1$ cavity vertices is completely random, the typical length $d(i,j)$ of a shortest-distance path between two cavity vertices $j$ and $k$ in the cavity graph ${\cal G}_K(N,K-1)$ is a large value. Actually, it can easily be shown \cite{Bollobas-1985} that this distance scales logarithmically with the cavity graph size $N$:
\begin{equation}
  \label{eq:lengthscale}
  d(j,k) \sim \ln N \ .
\end{equation}
Denote $\rho_{j,k,\ldots}(\sigma_j, \sigma_k, \ldots)$ as the joint probability distribution of the spin values  $\sigma_{j}, \sigma_{k},  \ldots$  of  a group of  cavity vertices in a cavity network. In a large random graph, according to Eq.~(\ref{eq:lengthscale}) the shortest path length between any two randomly chosen vertices is long. Therefore, as the zeroth-order approximation, one may assume that this joint probability distribution can be written as the following factorized form
\begin{equation}
  \label{eq:factorization}
\rho_{j,k,\ldots} (\sigma_{j}, \sigma_{k},  \ldots)
=\rho_j  (\sigma_j) \rho_k  (\sigma_k) \ldots   \ ,
\end{equation}
where $\rho_j(\sigma_j)$ is vertex $j$'s marginal spin value distribution as given by
Eq.~(\ref{eq:CavityField01}). Equation (\ref{eq:factorization}) is called the Bethe-Peierls approximation \cite{Bethe-1935,Peierls-1936a,Peierls-1936} in the literature. It assumes statistical independence among the spin states of the cavity vertices within a macroscopic state $\alpha$. (For recent references on extensions of the Bethe-Peierls approximation, see Refs.~\cite{Montanari-Rizzo-2005,Parisi-Slanina-2006,Chertkov-Chernyak-2006a,Chertkov-Chernyak-2006b}.)

\subsection{Cavity field distribution among different macroscopic states}
\label{subsec:field-distribution}

The cavity magnetization $m_j$ of the cavity vertex $j$ depends on the identity of the macroscopic state $\alpha$. Its value may be different in different macroscopic states. Let us denote ${\cal P}_j(m_j)$ as the fraction of macroscopic states in
which the cavity magnetization of vertex $j$ takes the value $m_j$, and denote ${\cal P}_{j,k,\ldots}(m_j, m_k, \ldots )$ as the fraction of macroscopic states in which the cavity magnetization of vertex $j, k, \ldots$ take the value $m_j, m_k, \ldots$, respectively. We extend the Bethe-Peierls approximation Eq.~(\ref{eq:factorization}) to the level of macroscopic states and assume that
\begin{equation}
  \label{eq:factorization-magnetization}
{\cal P}_{j,k,\ldots}(m_j, m_k , \ldots ) =
{\cal P}_j(m_j) {\cal P}_k (m_k) \ldots  \ .
\end{equation}
Equation~(\ref{eq:factorization-magnetization}) is equivalent to saying that, the fluctuations (among all the macroscopic states) of the cavity magnetization of two different cavity vertices are mutually independent of each other. Due to the absence of short loops in a random graph, Eq.~(\ref{eq:factorization-magnetization}) turns out to be a rather good approximation.

To obtain a self-consistent equation for the marginal probabilities $\{ {\cal P}_j(m_j) \}$, we add a new vertex $i$ and connect it to the  $K-1$  cavity vertices of  ${\cal G}_K(N,K-1)$. The quenched coupling constant $J_{i j}$ of each newly added edge is set to be $J_{i j}= \pm J$ with equal probability. This results in a new cavity network ${\cal G}_K(N+1,1)$ of $N+1$ vertices and one single cavity vertex
$i$ (see Fig.~\ref{fig:cavity01m1}B). In the corresponding macroscopic state $\alpha$ of the new cavity system ${\cal G}_K(N+1,1)$, the cavity vertex $i$ feels a cavity field $h_i$.

To calculate the cavity field $h_i$ in the new cavity system, we first notice that the energy difference between the cavity network ${\cal G}_K(N+1,1)$ and the old cavity network ${\cal G}_K(N,K-1)$ is
\begin{equation}
  \label{eq:deltaE1}
\Delta H_1 = - \sum\limits_{j \in \partial^\prime  i} J_{i j}\sigma_i \sigma_{j} \ ,
\end{equation}
where $\partial^\prime i$ denotes the set of $K-1$ nearest-neighbors of vertex $i$ in the cavity graph ${\cal G}_K(N+1,1)$. In the macroscopic state $\alpha$, the partition function for the cavity system ${\cal G}_K(N+1,1)$ is
\begin{eqnarray}
Z_\alpha \bigl({\cal G}_K(N+1,1) \bigr) &=&
\sum\limits_{\sigma_i} \sum\limits_{\{ \sigma_1, \ldots, \sigma_N\}\in  \alpha }
e^{-\beta H\bigl({\cal G}_K(N,K-1)\bigr) - \beta \Delta H_1 } \label{eq:Zacv03m3}
\\
&=& e^{-\beta F_\alpha \bigl( {\cal G}_K(N,K-1)\bigr)} \sum\limits_{\sigma_i}\frac{\sum\limits_{\sigma_{j}: j\in \partial^\prime i} e^{\beta \sum_{j\in \partial^\prime i} h_{j} \sigma_{j} - \beta \Delta H_1} }{
\sum\limits_{ \sigma_j: j\in \partial^\prime i} e^{\beta \sum_{j\in \partial^\prime i} h_{j} \sigma_{j}} }  \label{eq:Zacv03m2}
\\
&=& 
e^{-\beta F_\alpha \bigl( {\cal G}_K(N,K-1)\bigr)} \prod\limits_{j\in \partial^\prime i}\biggl[\frac{ \cosh \beta J_{i j} }{ \cosh \beta u(J_{i j}, h_{j})} \biggr]
\sum\limits_{\sigma_i}  \exp\bigl( \beta h_i \sigma_i \bigr)
\label{eq:Zacv03m1} \\
&=&  e^{-\beta F_\alpha \bigl( {\cal G}_K(N,K-1)\bigr)} \prod\limits_{j\in \partial^\prime i} \biggl[\frac{ \cosh \beta J_{i j} }{ \cosh \beta u(J_{i j}, h_{j})} \biggr] ( 2 \cosh \beta h_{i} )  \ .
\label{eq:Zacv03}
\end{eqnarray}
In going from Eq.~(\ref{eq:Zacv03m3}) to Eq.~(\ref{eq:Zacv03m2}), 
we have used the Bethe-Peierls approximation Eq.~(\ref{eq:factorization}).
$F_\alpha\bigl( {\cal G}_K(N,K-1) \bigr)$ is the free energy of the cavity system
${\cal G}_K(N,K-1)$ in its macroscopic state $\alpha$. In Eq.~(\ref{eq:Zacv03m1}), the quantity $u(J_{i j}, h_j)$ is defined as
\begin{equation}
  \label{eq:u}
u(J_{i j}, h_j) = \frac{ 1 }{ \beta} {\rm atanh} \bigl( \tanh \beta h_j \tanh \beta J_{i j} \bigr) \ ;
\end{equation}
and the quantity $h_i$ is calculated according to
\begin{equation}
  \label{eq:hia}
  h_i = \sum\limits_{j\in \partial^\prime  i} u(J_{i j}, h_{j}) \ .
\end{equation}
From Eq.~(\ref{eq:Zacv03m1}) we know that $h_i$ as expressed  by Eq.~(\ref{eq:hia})
is just the cavity field felt by vertex $i$ in the cavity network ${\cal G}_K(N+1,1)$.

If we know all the cavity fields on the nearest-neighbors of
vertex $i$, then we obtain the cavity field on vertex $i$ through the iterative
equation (\ref{eq:hia}). Consequently, the magnetization of the cavity vertex $i$ in the macroscopic state $\alpha$ of the cavity system ${\cal G}_K(N+1,1)$ is calculated as
\begin{equation}
  \label{eq:mi}
  m_i = \tanh \beta h_i =
   \frac{
    \prod\limits_{j\in \partial^\prime i} \bigl[ 1 + v_{i j}  m_j \bigr]
    - \prod\limits_{j\in \partial^\prime i} \bigl[ 1 - v_{i j}  m_j \bigr] } {
    \prod\limits_{j\in \partial^\prime i} \bigl[ 1 + v_{i j}  m_j \bigr] +
    \prod\limits_{j\in \partial^\prime i} \bigl[ 1 - v_{i j}  m_j ] } \ ,
\end{equation}
where $v_{i j}$ is a shorthand notation for $\tanh \beta J_{i j}$, i.e.,
\begin{equation}
  \label{eq:vij}
  v_{i j} \equiv \tanh \beta J_{i j} \ .
\end{equation}
Equation (\ref{eq:mi}) is an iterative equation for the cavity magnetization $m_i$
{\em within one macroscopic state $\alpha$}. 
As we have emphasized in Sec.~\ref{subsec:single01}, the input
cavity magnetization $m_j$ in Eq.~(\ref{eq:mi})  may be different in different macroscopic states. As a consequence, the
cavity magnetization $m_i$ of vertex $i$ does not necessarily take the same value
in different macroscopic states. On the contrary, its value may fluctuate a lot among different macroscopic states of the new cavity system ${\cal G}_K(N+1,1)$.
The task is to obtain an expression for the  marginal distribution ${\cal P}_i(m_i)$ 
of $m_i$ among different macroscopic states. 

From Eq.~(\ref{eq:Zacv03}), we know that, after the addition of the vertex $i$,
the free energy difference $\Delta F_1$ between the macroscopic state $\alpha$ of the system ${\cal G}_K(N+1,1)$ and that of the system ${\cal G}_K(N,K-1)$ is
\begin{eqnarray}
\Delta F_1&\equiv & -\frac{ 1 }{ \beta} \sum\limits_{j\in \partial^\prime  i} \ln\biggl(
  \frac{  \cosh \beta J_{i j} }{ \cosh \beta u(J_{i j}, h_j) } \biggr)
  - \frac{1 }{ \beta} \ln\bigl( 2 \cosh \beta h_i \bigr) 
  \label{eq:deltaF101} \\
&=& \frac{1 }{ 2 \beta} \sum\limits_{j\in\partial^\prime  i} \ln\bigl( 1- v_{i j}^2 \bigr)  - \frac{ 1 }{ \beta} \ln\biggl(
  \prod\limits_{j\in \partial^\prime  i} \bigl[1+ v_{i j} m_j \bigr]
  + \prod\limits_{j\in \partial^\prime  i} \bigl[1- v_{i j} m_j \bigr] \biggr) \ .
  \label{eq:deltaF102}
\end{eqnarray}
According to the free energy Boltzmann distribution Eq.~(\ref{eq:BoltzmannDistribution02}), each macroscopic state is weighted by
the Boltzmann factor $\exp(-y F_\alpha)$. After the addition of vertex $i$, the total partition function of the system ${\cal G}_K(N+1,1)$ is
\begin{eqnarray}
{\cal Z}\bigl( {\cal G}_K(N+1,1) \bigr) &=& \sum\limits_{\alpha} Z_\alpha\bigl( {\cal G}_K(N,K-1) \bigr) \exp(-y \Delta F_1)  \label{eq:rho-m1a} \\
&=& \biggl[ \sum\limits_{\alpha} Z_\alpha\bigl( {\cal G}_K(N,K-1) \bigr)  \biggr] \prod\limits_{j\in \partial^\prime i} \biggl[ \int {\rm d} m_j {\cal P}_j(m_j) \biggr] \exp(-y \Delta F_1) \ .
\label{eq:rho-m1}
\end{eqnarray}
We have used the factorization approximation Eq.~(\ref{eq:factorization-magnetization}) in going from Eq.~(\ref{eq:rho-m1a}) to Eq.~(\ref{eq:rho-m1}). Similarly, the total weight of those macroscopic states of the system ${\cal G}_K(N+1,1)$ in which the cavity magnetization of vertex $i$ being equal to $m_i$ is expressed as
\begin{eqnarray}
 {\cal Z}\bigl( {\cal G}_K(N+1,1); m_i \bigr) &=&
\biggl[ \sum\limits_{\alpha} Z_\alpha\bigl( {\cal G}_K(N,K-1) \bigr)  \biggr] \times \nonumber \\
& &  \prod\limits_{j\in \partial^\prime i} \biggl[ \int {\rm d} m_j {\cal P}_j(m_j) \biggr] \exp(-y \Delta F_1) \delta\Biggl(m_i-\frac{
    \prod\limits_{j\in \partial^\prime i} \bigl[ 1 + v_{i j}  m_j \bigr]
   - \prod\limits_{j\in \partial^\prime i} \bigl[ 1 - v_{i j}  m_j \bigr] }
   {
   \prod\limits_{j\in \partial^\prime i} \bigl[ 1 + v_{i j}  m_j \bigr] +
   \prod\limits_{j\in \partial^\prime i} \bigl[ 1 - v_{i j}  m_j ] } \Biggr) \ .
\label{eq:rho-m2}
\end{eqnarray}
From Eq.~(\ref{eq:rho-m1}) and Eq.~(\ref{eq:rho-m2}) we realize that, in the new system, the fraction of macroscopic states in which vertex $i$ bearing a cavity magnetization $m_i$ is equal to
\begin{equation}
  \label{eq:rho}
{\cal P}_i (m_i) \equiv \frac{{\cal Z}\bigl( {\cal G}_K(N+1,1); m_i \bigr)}{{\cal Z}\bigl( {\cal G}_K(N+1,1) \bigr)} \propto 
\prod\limits_{j\in \partial^\prime  i} \biggl[ \int {\rm d} m_j {\cal P}_j(m_j) \biggr]  e^{-y \Delta F_1} \delta\Biggl( m_i -
\frac{
    \prod\limits_{j\in \partial^\prime i} \bigl[ 1 + v_{i j}  m_j \bigr]
   - \prod\limits_{j\in \partial^\prime i} \bigl[ 1 - v_{i j}  m_j \bigr] }
   {
   \prod\limits_{j\in \partial^\prime i} \bigl[ 1 + v_{i j}  m_j \bigr] +
   \prod\limits_{j\in \partial^\prime i} \bigl[ 1 - v_{i j}  m_j ] } \Biggr) 
 \ .
\end{equation}

Equation (\ref{eq:rho}) is a self-consistent iterative equation for the cavity distributions $\{ {\cal P}_i\}$.  A steady state solution of Eq.~(\ref{eq:rho}) can be obtained by population dynamics \cite{Mezard-Parisi-2001,Mezard-Zecchina-2002}. An array of ${\cal N}$ probability distributions ${\cal P}_i$ are stored.  At each step of the population dynamics, $K-1$ probability distributions are randomly chosen from this array of ${\cal N}$ stored distributions, and a new probability distribution is generated by using Eq.~(\ref{eq:rho}). This new distribution then
replaces a randomly chosen old probability distribution in the array. This iteration process is repeated many times until the population dynamics reaches a steady state.

\section{Grand free energy density}
\label{sec:densities}

\begin{figure}[ht]
  \begin{center}
    \includegraphics[width=0.25\linewidth]{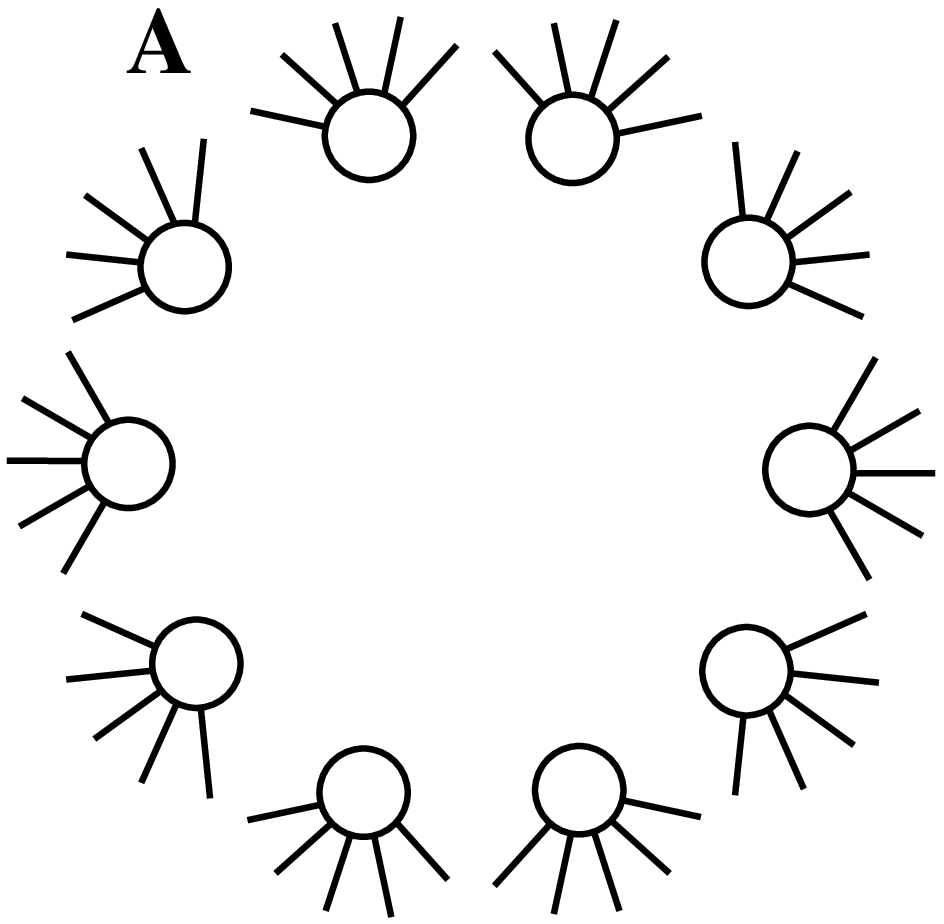}
    \includegraphics[width=0.25\linewidth]{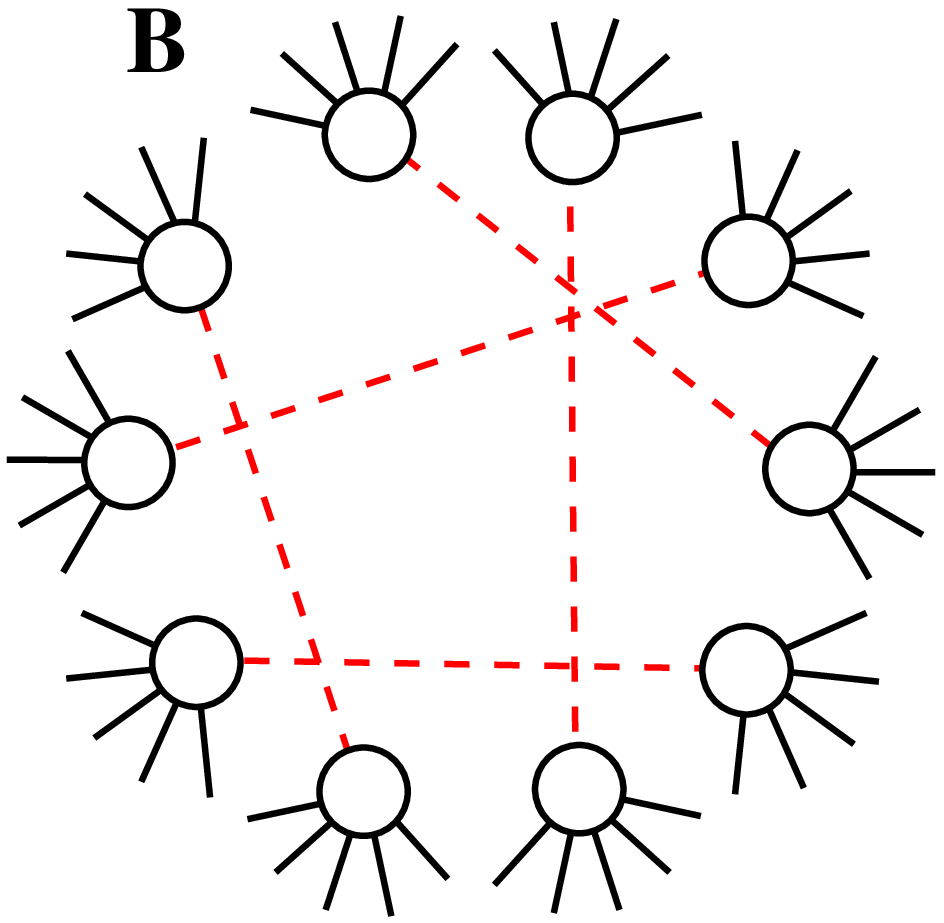}
    \includegraphics[width=0.25\linewidth]{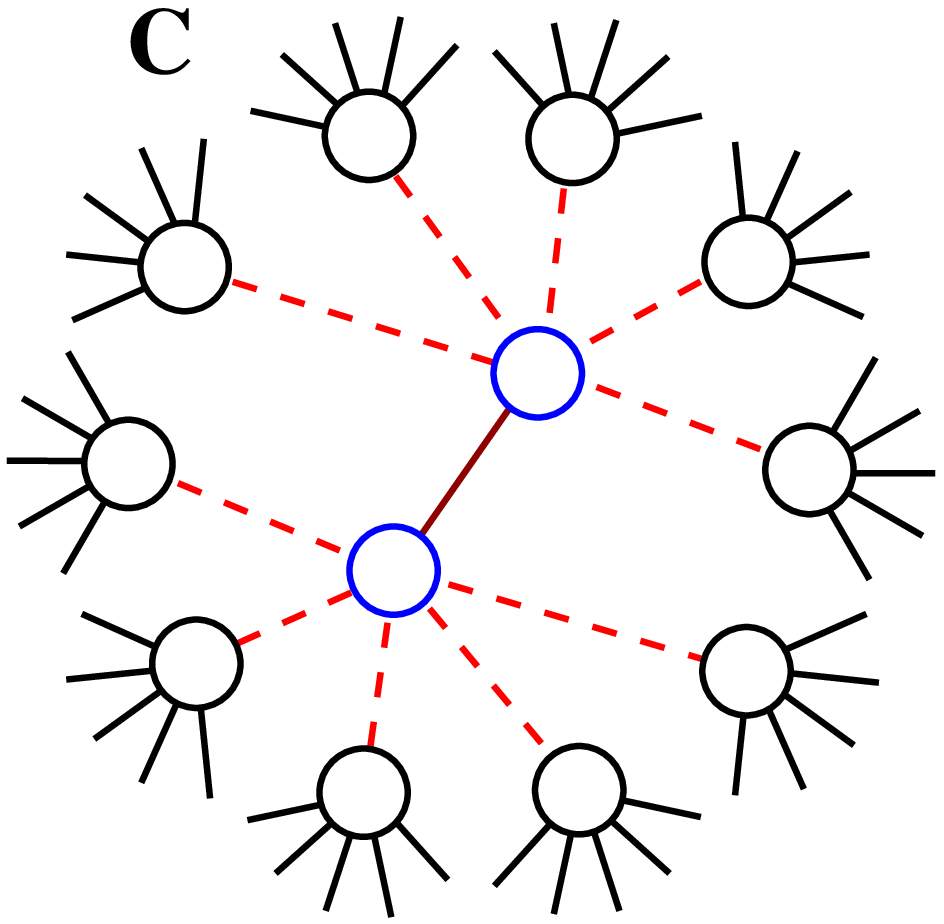}
  \end{center}
  \caption{\label{fig:cavity01}
The cavity approach to the spin-glass model Eq.~(\ref{eq:energy01}) on a random regular graph of connectivity $K = 6$. (A): part of a cavity graph with $2 (K-1)$ cavity vertices. (B) and (C): construction of a random regular graph of size $N$ (B) and $N+2$ (C) from the  cavity  graph with $N$ vertices and $2 (K-1)$ cavity vertices.
}
\end{figure}

The free energy of a macroscopic state $\alpha$ is defined formally by Eq.~(\ref{eq:Falpha}). As we noted before, this expression is not directly applicable since we do not know what are the microscopic configurations of state $\alpha$. The cavity approach \cite{Mezard-Parisi-2001} circumvents this problem by calculating the grand free energy difference between a system of $N$ vertices and an enlarged system of $N+2$ vertices. As demonstrated in Fig.~\ref{fig:cavity01}, a random network ${\cal G}_K(N)$ can be constructed from a random cavity network ${\cal G}_K\bigl(N, 2(K-1)\bigr)$ by adding $K-1$ new edges. The grand free energy difference between these two systems can be calculated. Similarly, one can add two new vertices and $2 K - 1$ new edges to change the same random cavity network ${\cal G}_K\bigl(N, 2(K-1) \bigr)$ into a random regular network ${\cal G}_K(N+2)$. The grand free energy difference between these two systems can also be calculated. From the two grand free energy differences, one can obtain the grand free energy density $g(\beta; y)$ for a random regular system ${\cal G}_K(N)$. The mean free energy density of a macroscopic state and the complexity of the system can then be obtained from $g(\beta; y)$.

\subsection{From the cavity network ${\cal G}_K\bigl(N,2 (K-1)\bigr)$ to the regular network ${\cal G}_K(N)$}
\label{subsec:NtoN}

The difference $\Delta H_2$ between the configurational energy of the system on
the random graph ${\cal G}_K(N)$ in Fig.~\ref{fig:cavity01}B and that of the system  on the random cavity graph ${\cal G}_K\bigl(N, 2 (K-1) \bigr)$ in Fig.~\ref{fig:cavity01}A is
\begin{equation}
  \label{eq:EnergyDifference02}
  \Delta H_2 = - \sum\limits_{(i,j)\in e_1} J_{i j}  \sigma_i \sigma_j \ ,
\end{equation}
where $e_1$ denotes the set of $K-1$ newly added edges 
in going from ${\cal G}_K\bigl(N, 2 (K-1) \bigr)$ to ${\cal G}_K(N)$. 
Each edge in set $e_1$
connects two cavity vertices of ${\cal G}_K\bigl(N, 2(K-1) \bigr)$.
Following the analytical procedure of Sec.~\ref{subsec:field-distribution}, we know that difference between the free energy of a macroscopic state $\alpha$ of the system ${\cal G}_K(N)$ and that of the same macroscopic state $\alpha$ of the cavity system ${\cal G}_K\bigl(N, 2(K-1)\bigr)$ is
\begin{equation}
  \label{eq:FreeEnergy01}
  \Delta F_2 =  \sum\limits_{(i,j)\in e_1} \Delta F^{(i,j)} \ ,
\end{equation}
where
\begin{equation}
  \label{eq:FreeEnergy02}
  \Delta F^{(i,j)} = \frac{ 1 }{ 2 \beta} \ln \bigl(1 - v_{i j}^2 \bigr)
- \frac{ 1 }{ \beta}  \ln \bigl( 1 + v_{i j} m_i m_j \bigr)
\end{equation}
can be understood as the free energy increase caused by the addition of an edge (with coupling $J_{i j}$) between two cavity vertices $i$ and $j$.

The grand free energy $G(N, \beta; y)$ of the new system ${\cal G}_K(N)$ is related to the grand free energy $G^{({\rm cv})}(N,\beta;y)$ of the old cavity system ${\cal G}_K\bigl(N,2(K-1)\bigr)$ by the following equation
\begin{equation}
 \label{eq:gfd-i}
G(N, \beta; y) =  G^{({\rm cv})}(N, \beta; y) + \sum\limits_{(i,j)\in e_1} 
\Delta G^{(i,j)} \ ,
\end{equation}
where
\begin{equation}
\label{eq:dGij}
\Delta G^{(i,j)} = -\frac{1}{y} \ln\Bigl[\int {\rm d} m_i \int {\rm d} m_j {\cal P}_i(m_i) {\cal P}_j (m_j)  \exp\bigl(-y \Delta  F^{(i,j)} \bigr) \Bigr] \ 
\end{equation}
is the increase to the grand free energy caused by adding an edge $(i,j)$.

\subsection{From the cavity network ${\cal G}_K\bigl(N, 2 (K-1)\bigr)$ to the regular network ${\cal G}_K(N+2)$}
\label{subsec:NtoNplus2}

The random network ${\cal G}_K(N+2)$ in Fig.~\ref{fig:cavity01}C is constructed from the random cavity network ${\cal G}_K\bigl(N, 2 (K-1) \bigr)$ of Fig.~\ref{fig:cavity01}A by adding two vertices ($i_0$ and $j_0$). Vertex $i_0$ is connected to $K-1$ cavity vertices ($j_1, \ldots, j_{K-1}$) of ${\cal G}_K\bigl(N, 2(K-1) \bigr)$, and vertex $j_0$ is connected to the remaining $K-1$ cavity vertices ($i_1, \ldots, i_{K-1}$) of ${\cal G}_K\bigl(N,2(K-1)\bigr)$. Vertex $i_0$ and $j_0$ are directly connected by a new edge $(i_0,j_0)$, so that every vertex in the graph ${\cal G}_K(N+2)$ has degree $K$.
The energy difference between the system ${\cal G}_K(N+2)$ and the cavity system ${\cal G}_K\bigl(N,2(K-1)\bigr)$ is
\begin{equation}
  \label{eq:EnergyDifference06}
\Delta H_3 = - J_{i_0 j_0 }  \sigma_{i_0} \sigma_{j_0} - \sum\limits_{s=1}^{K-1} J_{i_0 j_s} \sigma_{i_0} \sigma_{j_s}- \sum\limits_{s=1}^{K-1} J_{j_0  i_s} \sigma_{j_0} \sigma_{i_s} \ .
\end{equation}

The increase $\Delta F_3$ in the free energy of macroscopic state $\alpha$ due to the addition of two new vertices and $2 K -1 $ new edges can be obtained following the same procedure as given in Sec.~\ref{subsec:field-distribution}.  We find that 
\begin{equation}
  \Delta F_3 =  \Delta F^{(i_0)} + \Delta F^{(j_0)} - \Delta F^{(i_0,j_0)} \ .
  \label{eq:FreeEnergy03}
\end{equation}
In Eq.~(\ref{eq:FreeEnergy03}), $\Delta F^{(i_0)}$ is the free energy increase caused by adding vertex $i_0$ and connecting it to the set $\partial i_0=\{j_0, j_1,\ldots, j_{K-1} \}$ of $K$ vertices. The explicit expression for $\Delta F^{(i_0)}$ is
\begin{equation}
\Delta F^{(i_0)} =
\frac{1 }{ 2 \beta} \sum\limits_{j\in\partial i_0} \ln\bigl( 1- v_{i_0 j}^2 \bigr)  - \frac{ 1 }{ \beta} \ln\biggl(
  \prod\limits_{j\in \partial  i_0} \bigl[1+ v_{i_0 j} m_j \bigr]
  + \prod\limits_{j\in \partial i_0} \bigl[1- v_{i_0 j} m_j \bigr] \biggr) \ .
  \label{eq:deltaFi0}
\end{equation}
The expression for $\Delta F^{(j_0)}$ has the same form as Eq.~(\ref{eq:deltaFi0}). Let us emphasize that, in Eq.~(\ref{eq:deltaFi0}), $m_{j_0}$ is the cavity magnetization of vertex $j_0$ when vertex $i_0$ is not added, i.e.,
\begin{eqnarray}
m_{i_0} &=&  \frac{
\prod\limits_{j_s \in \partial i_0 \backslash j_0} \bigl[ 1 + v_{i_0 j_s}  m_{j_s} \bigr]- \prod\limits_{j_s\in \partial i_0 \backslash j_0} \bigl[ 1 - v_{i_0 j_s }  m_{j_s} \bigr] }{\prod\limits_{j_s\in \partial i_0 \backslash j_0} \bigl[ 1 + v_{i_0 j_s}  m_{j_s} \bigr] + \prod\limits_{j_s\in \partial i_0 \backslash j_0} \bigl[ 1 - v_{i_0 j_s }  m_{j_s} \bigr] } \ ,
  \label{eq:m0val}  \\
m_{j_0} &=& \frac{
\prod\limits_{i_s \in \partial j_0 \backslash i_0} \bigl[ 1 + v_{j_0  i_s}  m_{i_s} \bigr]- \prod\limits_{i_s \in \partial j_0  \backslash i_0 } \bigl[ 1 - v_{j_0 i_s}  m_{i_s } \bigr] }{\prod\limits_{i_s \in \partial j_0 \backslash i_0} \bigl[ 1 + v_{j_0  i_s}  m_{i_s} \bigr] + \prod\limits_{i_s\in \partial j_0 \backslash i_0 } \bigl[ 1 - v_{j_0  i_s}  m_{i_s} \bigr] } \ .
\label{eq:m0pval}
\end{eqnarray}
The term $\Delta F^{(i_0,j_0)}$ of Eq.~(\ref{eq:FreeEnergy03}) is the free energy increase caused by setting up an edge between vertex $i_0$ and vertex $j_0$; its expression is given by Eq.~(\ref{eq:FreeEnergy02}), with $m_{i_0}$ and $m_{j_0}$ being determined by Eq.~(\ref{eq:m0val}) and Eq.~(\ref{eq:m0pval}), respectively.
The free energy increase Eq.~(\ref{eq:FreeEnergy03}) can be intuitively understood as follows: Since the contribution of the edge $(i_0, j_0)$ is counted twice in
$\Delta F^{(i_0)}$ and $\Delta F^{(j_0)}$, the free energy increase should be corrected with an edge term.

With these preparations, we can calculate the total grand free energy $G(N+2, \beta; y)$ of the system ${\cal G}_K(N+2)$. Similar to Eq.~(\ref{eq:gfd-i}), we find that
\begin{equation}
\label{eq:delta-G-2-01}
G(N+2, \beta; y) = G^{({\rm cv})}(N, \beta; y) + \Delta G^{(i_0)} + \Delta G^{(j_0)} - \Delta G^{(i_0, j_0)} \ .
\end{equation}
In Eq.~(\ref{eq:delta-G-2-01}), $\Delta G^{(i_0)}$ is the increase to the grand free energy caused by adding vertex $i_0$ and connecting it to $K$ vertices, with
\begin{equation}
\label{eq:dGi0}
\Delta G^{(i_0)} = -\frac{1}{y} \ln\Biggl( \int \prod\limits_{j_s\in \partial i_0}  {\rm d} m_{j_s} {\cal P}_{j_s}(m_{j_s}) \exp(-y \Delta F^{(i_0)}) \Biggr) \ ,
\end{equation}
where ${\cal P}_{j_0}(m_{j_0})$ is the cavity magnetization distribution of vertex $j_0$ in the absence of vertex $i_0$. $\Delta G^{(i_0, j_0)}$ in Eq.~(\ref{eq:delta-G-2-01}) is calculated through Eq.~(\ref{eq:dGij}) using ${\cal P}_{i_0}$ and ${\cal P}_{j_0}$.

\subsection{Averaging over the quenched randomness}
\label{subsec:averaging}

The grand free energy density of the spin-glass system is
\begin{equation}
\label{gfd-unaveraged}
g(\beta;y)=\lim\limits_{N\to \infty} \frac{ G(N+2,\beta; y) - G(N,\beta;y)}{2} \ .
\end{equation}
An explicit expression for $g(\beta; y)$ can be written down by applying Eq.~(\ref{eq:gfd-i}) and Eq.~(\ref{eq:delta-G-2-01}), which is a function of the quenched randomness in the system. The grand free energy density has the nice property of self-averaging \cite{Mezard-etal-1987}, namely the value of $g(\beta;y)$ as calculated for a typical system is equal to the averaged value of $g(\beta;y)$ over many systems with different realizations of the quenched randomness in the graph connection pattern and in the edge coupling constants. When the quenched randomness is averaged out, we obtain that
\begin{equation}
 \label{eq:g-final}
g(\beta; y) = \overline{ \Delta G^{(i)} } - \frac{K}{2} \overline{ \Delta G^{(i,j)} } \ ,
\end{equation}
where $\Delta G^{(i)}$ and $\Delta G^{(i,j)}$ are calculated through Eq.~(\ref{eq:dGi0}) and Eq.~(\ref{eq:dGij}), respectively; and  an overline indicates averaging over the quenched randomness of the spin-glass system.

The mean free energy density of a macroscopic state of the system is related to $g(\beta; y)$ by
\begin{equation}
 \label{eq:f-final}
f(\beta; y) = \frac{\partial y g(\beta; y)}{\partial y}  = g(\beta; y) + y \frac{ \partial g(\beta; y)}{\partial y} \ ,
\end{equation}
and the complexity of the system at given value of the reweighting parameter $y$ is
\begin{equation}
 \label{eq:sigma-final}
\Sigma( \beta; y) \equiv y\Bigl( f-g(\beta; y)\Bigr) =y^2 \frac{\partial g(\beta; y)}{\partial y} \ .
\end{equation}

\section{thermodynamic quantities for a single instance of the system}
\label{sec:single-instance}

The discussion in Sec.~\ref{sec:densities} was concerned with the typical properties
of an ensemble of spin-glass systems governed by a given distribution of quenched
randomness. One important advantage of the cavity approach is that, for a single instance of the quenched randomness, the thermodynamic properties can also be calculated. Under the Bethe-Peierls approximation, the total grand free energy of
a spin-glass system Eq.~(\ref{eq:energy01}) on a graph ${\cal G}$ with couplings $\{J_{i j} \}$ is equal to
\begin{equation}
 \label{eq:gfe-single-01}
G_{ \{ J_{i j} \}, {\cal G} }(\beta; y ) = \sum\limits_{i \in {\cal G}}
\Delta G_{ \{ J_{i j} \}, {\cal G} }^{(i)}(\beta; y) - \sum\limits_{(i,j)\in {\cal G}} \Delta G_{\{ J_{i j} \}, {\cal G} }^{(i j)}(\beta; y) \ .
\end{equation}
It is easy to see that the above equation is consistent with Eq.~(\ref{eq:g-final}).
In Eq.~(\ref{eq:gfe-single-01}), $\Delta G_{ \{ J_{i j} \}, {\cal G} }^{(i)}$ denotes the contribution of vertex $i$ and its associated edges to the total grand free energy of the system; its expression has the same form as Eq.~(\ref{eq:dGi0}):
\begin{equation}
 \label{eq:gfe-single-02}
\Delta G_{ \{ J_{i j} \}, {\cal G} }^{(i)}(\beta; y)=
-\frac{1}{y}\ln\biggl(\prod\limits_{j\in \partial i} \Bigl[\int {\rm d} m_{j\to i} {\cal P}_{j\to i}(m_{j\to i}) \Bigr]
e^{-y \Delta F^{(i)} } \biggr) \ ,
\end{equation}
where $m_{j\to i}$ is the cavity magnetization of vertex $j$ (with respect to vertex $i$) in a macroscopic state $\alpha$, and ${\cal P}_{j\to i}(m_{j\to i})$ is the probability distribution of this cavity magnetization among all the macroscopic states; $\Delta F^{(i)}$ is the free energy contribution (in a given macroscopic state) of vertex $i$ and its associated edges, which has the same form as Eq.~(\ref{eq:deltaFi0}) but with $m_j$ replaced by $m_{j\to i}$. Similarly, the term $\Delta G_{\{ J_{i j} \}, {\cal G} }^{(i j)}$ in Eq.~(\ref{eq:gfe-single-01}) is the contribution of an edge $(i,j)$ to the total grand free energy of the system; it is expressed as
\begin{equation}
 \label{eq:gfe-single-03}
\Delta G_{\{ J_{i j} \}, {\cal G} }^{(i j)}(\beta; y)=
-\frac{1}{y}\ln\biggl(\int {\rm d} m_{j\to i} \int {\rm d}
m_{i\to j} {\cal P}_{j\to i}(m_{j\to i})
{\cal P}_{i\to j}(m_{i\to j}) e^{-y \Delta F^{(i j)}} \biggr) \ ,
\end{equation}
where $\Delta F^{(i j)}$ is given by Eq.~(\ref{eq:FreeEnergy02}) but with
$m_i$ replaced by $m_{i\to j}$ and $m_j$ replaced by $m_{j\to i}$.

The set of $ 2 M$ ($M$ being the total number of edges in the graph ${\cal G}$) probability distributions ${\cal P}_{j\to i}(m_{j\to i})$ in Eq.~(\ref{eq:gfe-single-01}) should be carefully chosen such that the total grand free energy achieves a minimal value at given $(\beta, y)$ values. In other words, for any edge $(i,j)$ of the graph ${\cal G}$, the variation of $G_{\{J_{i j}\},{\cal G}}$ with respect to both ${\cal P}_{j\to i}$ and ${\cal P}_{i\to j}$ should vanish:
\begin{equation}
 \label{eq:gfe-single-04}
\frac{\delta G_{\{J_{i j} \}, {\cal G} } }{\delta {\cal P}_{j\to i} }
=\frac{\delta G_{\{J_{i j} \}, {\cal G} } }{\delta {\cal P}_{i\to j} }
= 0 \ .
\end{equation}
Equation (\ref{eq:gfe-single-04}) results in the following self-consistent Bethe-Peierls equation for the $P_{i \to j}$'s:
\begin{equation}
 \label{eq:gfe-single-05}
{\cal P}_{i\to j}(m_{i\to j}) \propto \prod\limits_{k\in \partial i\backslash j} \biggl[ \int {\rm d} m_{k\to i} {\cal P}_{k\to i}(m_{k\to i}) \biggr] e^{-y \Delta F_j^{(i)}} \delta\Biggl( m_{i\to j}-\frac{ \prod\limits_{k\in \partial i \backslash j} [1 + v_{i k}  m_{k\to i} ] - \prod\limits_{k\in \partial i \backslash j} [1 - v_{i k}  m_{k\to i} ]}
{\prod\limits_{k\in \partial i\backslash j} [1 + v_{i k}  m_{k\to i}]+
   \prod\limits_{k\in \partial i\backslash j} [1 - v_{i k}  m_{k\to i}] }
\Biggr) \ ,
\end{equation}
where
\begin{equation}
 \label{eq:gfe-single-06}
\Delta F_j^{(i)}=
\frac{1 }{ 2 \beta} \sum\limits_{k\in\partial i\backslash j} \ln\bigl( 1- v_{i k}^2 \bigr)- \frac{ 1 }{ \beta} \ln\biggl(
\prod\limits_{k\in \partial i\backslash j} \bigl[1+ v_{i k} m_{k\to i} \bigr]
+\prod\limits_{k\in \partial i\backslash j} \bigl[1- v_{i k} m_{k\to i} \bigr] \biggr) \ .
\end{equation}
$\Delta F_j^{(i)}$ has the same physical meaning as $\Delta F_1$ of Sec.~\ref{subsec:field-distribution}. Equation (\ref{eq:gfe-single-06}) is consistent with Eq.~(\ref{eq:rho}) of Sec.~\ref{subsec:field-distribution}.

Similar to Eq.~(\ref{eq:f-final}) and Eq.~(\ref{eq:sigma-final}), the mean free energy of a macroscopic state of the sample is expressed as
\begin{equation}
 \label{eq:fe-single-02}
F_{\{ J_{i j}\}, {\cal G}}(\beta; y)= G_{\{J_{i j} \},{\cal G}}(\beta; y)
+ y \frac{\partial G_{\{J_{i j} \},{\cal G}}(\beta; y)}{\partial y} \ ,
\end{equation}
and the complexity of the system is calculated through
\begin{equation}
 \label{eq:fe-single-03}
\Sigma_{\{ J_{i j}\}, {\cal G}}(\beta; y)=\frac{1}{N} y^2
\frac{\partial G_{\{J_{i j} \},{\cal G}}(\beta; y)}{\partial y} \ .
\end{equation}

Because of Eq.~(\ref{eq:gfe-single-04}), the first derivative of $G_{\{J_{i j}\}, {\cal G}}$ with respect to $y$ can easily be expressed. We find that Eq.~(\ref{eq:fe-single-02}) can be re-written as
\begin{equation}
 \label{eq:free-energy-1RSB}
F_{\{ J_{i j}\}, {\cal G}}(\beta; y)=\sum\limits_{i \in {\cal G}}
\bigl\langle \Delta F^{(i)}\bigr\rangle- \sum\limits_{(i,j)\in {\cal G}} \bigl\langle \Delta F^{(i j)} \bigr\rangle\ ,
\end{equation}
where
\begin{eqnarray}
\bigl\langle \Delta F^{(i)} \bigr\rangle &=&\frac{\prod\limits_{j\in \partial i} \Bigl[\int {\rm d} m_{j\to i} {\cal P}_{j\to i}(m_{j\to i}) \Bigr] e^{-y \Delta F^{(i)} } \Delta F^{(i)} }{ \prod\limits_{j\in \partial i} \Bigl[\int {\rm d} m_{j\to i} {\cal P}_{j\to i}(m_{j\to i}) \Bigr] e^{-y \Delta F^{(i)} } } \ ,
\label{eq:average-dfi} \\
\bigl\langle \Delta F^{(i j)} \bigr\rangle &=&\frac{\int {\rm d} m_{j\to i} \int {\rm d} m_{i\to j} {\cal P}_{j\to i}(m_{j\to i}) {\cal P}_{i\to j}(m_{i\to j}) e^{-y \Delta F^{(i j)}} \Delta F^{(i j)} }{\int {\rm d} m_{j\to i} \int {\rm d} m_{i\to j} {\cal P}_{j\to i}(m_{j\to i}){\cal P}_{i\to j}(m_{i\to j}) e^{-y \Delta F^{(i j)}} } \ .
\label{eq:average-dfij}
\end{eqnarray}
The mean free energy of a macroscopic state is also decomposed into vertex contributions and edge contributions.

According to the discussion in Sec.~\ref{sec:mfep}, in the mean-field theory, the reweighting parameter $y$ should be chosen such that its value is closest to the physical inverse temperature $\beta$. Therefore, at a given value of $\beta$, the mean free energy $F_{\{J_{i j}\}, {\cal G} }(\beta)$ of the system is obtained by setting $y=\beta$ in Eq.~(\ref{eq:fe-single-02}) or Eq.~(\ref{eq:free-energy-1RSB}), provided that the complexity $\Sigma_{{\cal G},  \{ J_{i j}\}}(\beta; y)$ is always non-negative in the range of $0 \leq y \leq \beta$. If, on the other hand,  $\Sigma_{{\cal G},  \{ J_{i j}\}}(\beta; y)$ changes from being positive to being negative at a point of $y=y^* < \beta$, we should set $y$ to $y=y^*$. In this later situation, the mean free energy of the system  is equal to the maximal value of the grand free energy, $G_{{\cal G},\{J_{i j} \}}(\beta; y^*)$. After the mean free energy  $F_{\{J_{i j}\}, {\cal G} }(\beta)$ of the spin-glass system is obtained, the mean energy of a macroscopic state is calculated through
\begin{equation}
E_{\{J_{i j}\},{\cal G}}(\beta)= F_{\{J_{i j}\}, {\cal G} }(\beta)+ \beta \frac{ {\rm d} F_{\{J_{i j}\}, {\cal G} }(\beta) }{{\rm d} \beta} \ ,
\label{eq:mean-energy}
\end{equation}
and the mean entropy of a macroscopic state is calculated through
\begin{equation}
 \label{eq:mean-entropy}
S_{\{J_{i j}\},{\cal G}}(\beta)= \beta^2 \frac{ {\rm d} F_{\{J_{i j}\}, {\cal G} }(\beta) }{{\rm d} \beta} \ .
\end{equation}

Ensemble-averaged densities of mean energy and mean entropy can also be obtained from Eqs.~(\ref{eq:mean-energy}) and (\ref{eq:mean-entropy}) by averaging out the quenched randomness. An alternative derivation for the mean energy density is given in the appendix.

\section{Numerical results for the $\pm J$ model with $K=6$}
\label{sec:numerics}

We have applied the mean-field treatments of Sec.~\ref{sec:cavity-field-distribution} and Sec.~\ref{sec:densities} to the $\pm J$ spin-glass model Eq.~(\ref{eq:energy01}) on a random regular graph of vertex degree $K=6$. (In what follows, energy is in units of $J$.) At temperature $T=0.8$ (i.e., $\beta=1.25$), Carrus, Marinari and Zuliani (as cited in Ref.~\cite{Mezard-Parisi-2001}) estimated that the energy density of this system is $\epsilon = -1.799 \pm 0.001$. The mean-field theory of Ref.~\cite{Mezard-Parisi-2001} also predicted the same result. This reference also predicted a mean-free energy density of $f= -1.858 \pm 0.002$.

In the present population dynamics simulation, each probability profile ${\cal P}_i(m_i)$ in Eq.~(\ref{eq:rho}) was represented by an vector of $m=200$ elements. A total number of ${\cal N}$ such probability profiles were stored in the computer and they were updated using Eq.~(\ref{eq:rho}). In our work, various ${\cal N}$ values ranging from $32$ to $6,000$ were used. In each elementary update of the probability distribution ${\cal P}_i(m_i)$, a Metropolis importance sampling \cite{Newman-Barkema-1999} with inverse temperature $y$ was exploited to generate a total number of $n \times m$ magnetizations $m_i$. These magnetizations were stored in an ordered set. The averaged value of the first (second, third, $\ldots$) $n$ magnetizations in this set is assigned to the first
(second, third,$\ldots$) element of the new ${\cal P}_i$. We chose $n=200$ in our simulations.

\begin{figure}[ht]
 \centerline{\includegraphics[width=0.5\linewidth]{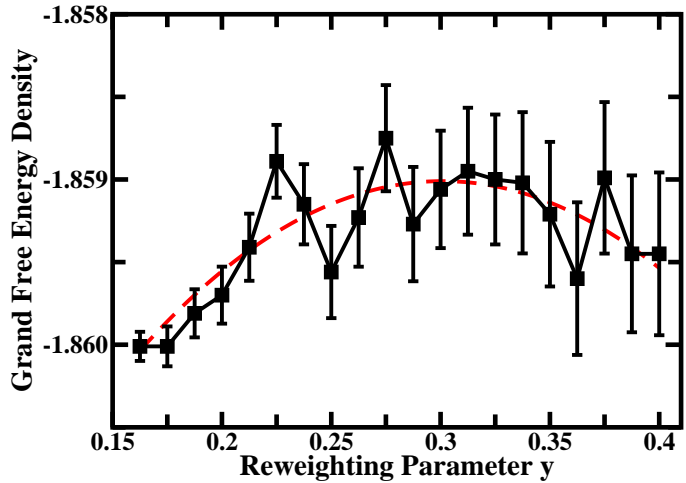} }
\caption{\label{fig:gy}
The grand free energy density $g(\beta;y)$ of the $\pm J$ spin-glass model on a random regular graph of vertex degree $K=6$. The inverse temperature is fixed to $\beta=1.25$. Symbols are results obtained by population dynamics simulations using a population size of ${\cal N}=256$. The dashed line is a fit to the data using  $g(y)=a-b (y-y^*)^2$, with the fitting parameters $a=-1.8590\pm 0.0001$ and $y^*=0.30\pm 0.01$.
}
\end{figure}
\begin{figure}[ht]
\includegraphics[width=0.5\linewidth]{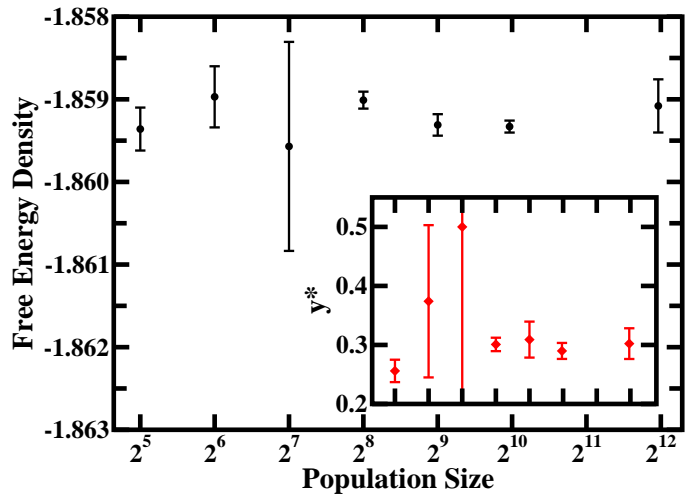}
 \caption{\label{fig:mfd}
Mean free energy density of the $\pm J$ spin-glass model on a random regular
graph of vertex degree $K=6$ at $\beta=1.25$ as obtained by population dynamics with different population size ${\cal N}$. (Inset) the optimal reweighting parameter $y^*$.
}
\end{figure}
\begin{figure}[ht]
\centerline{\includegraphics[width=0.6\linewidth,angle=270]{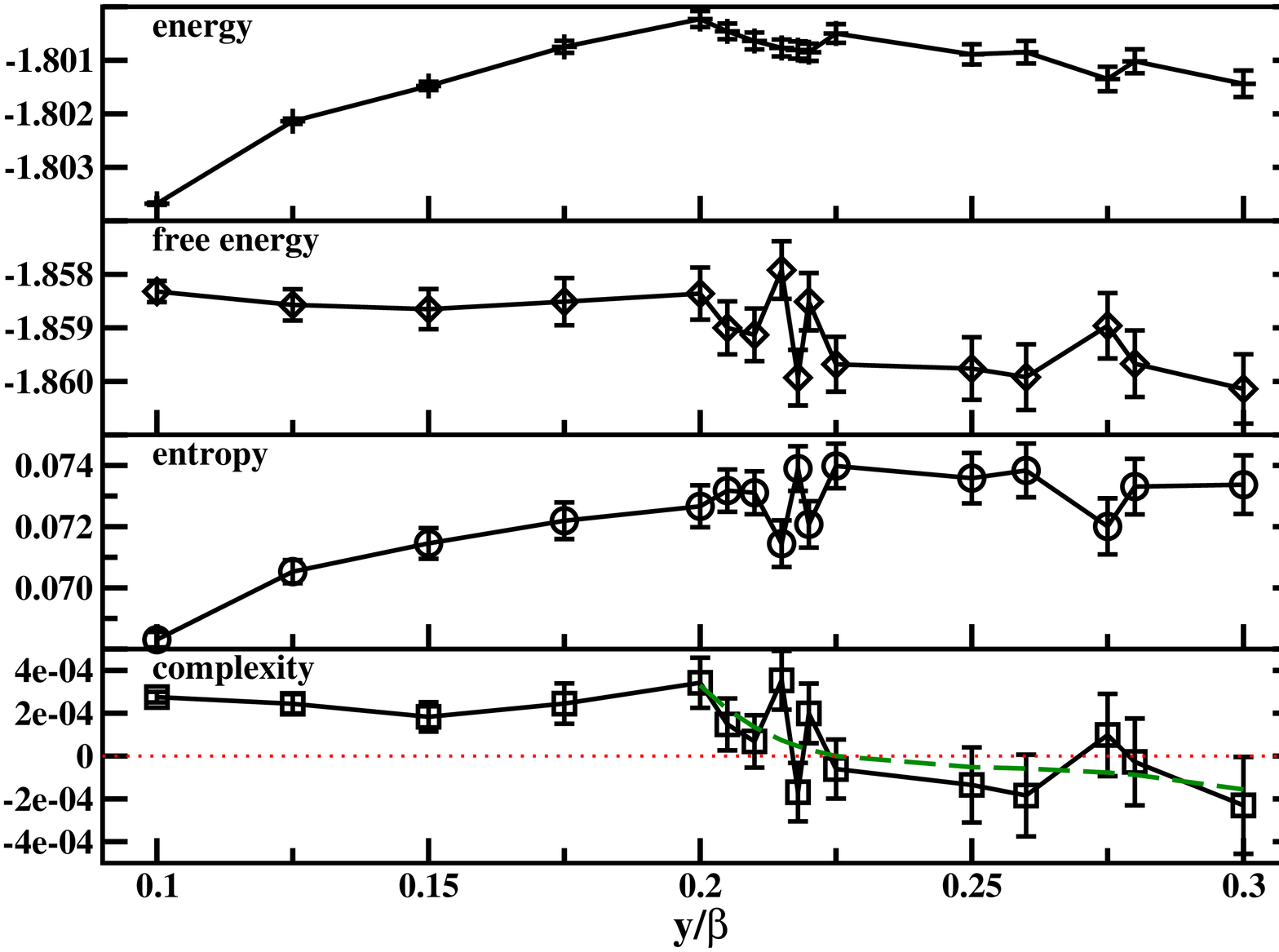}}
  \caption{ \label{fig:result01}
The energy density, free energy density, entropy density, and complexity
as a function of the reweighting parameter $y$ for the $\pm J$ spin-glass model
on a random regular graph of vertex degree $K=6$ at inverse temperature
$\beta=1.25$. Energy unit is $J$.  The symbols represent numerical population dynamics results using ${\cal N}=4000$. The dashed line in the bottom  panel is a fitting to the numerical data with $\Sigma= a_0 + a_1/y + a_2 /y^2 + a_3 / y^3$ in the range of $0.2 \beta \leq y \leq 0.3 \beta$. To reduce computational time, for the simulations reported in this figure, a faster but less precise numerical recipe was used to generate a new probability profile ${\cal P}_i(m_i)$ using Eq.~(\ref{eq:rho}). This might be the reason for the relatively large fluctuations of the mean densities with the reweighting parameter $y$.
}
\end{figure}

The grand free energy density $g(\beta;y)$ of the system as calculated according to Eq.~(\ref{eq:g-final}) is shown in Fig.~\ref{fig:gy}. The grand free energy density $g(y)$ first increases with the reweighting parameter $y$ till $y$ reaches $y^*=0.30\pm 0.01$, at which point $g(\beta; y)$ attains a maximal value of $g=-1.8590\pm 0.0001$. This maximal value of $g(\beta; y)$ corresponds to the best estimate of the mean free energy density of the system by the present mean-field method. It is in close agreement with the prediction of Ref.~\cite{Mezard-Parisi-2001}. We have checked that the estimated mean free energy density value is not sensitive to the population size ${\cal N}$ (see Fig.~\ref{fig:mfd}). We have also calculated the mean energy density and mean entropy density  for the model system at $\beta=1.25$ using a population size of ${\cal N}=4,000$, with the mean energy density being $\epsilon=-1.8007\pm 0.0002$ and the mean entropy density being $s=0.0732 \pm 0.0005$ (see Fig.~\ref{fig:result01}). These predictions are also in good agreement with the numerical results of Ref.~\cite{Mezard-Parisi-2001} and with the simulational work of Carrus, Marinari and Zuliani. These good agreements of the present theoretical results with earlier simulational and numerical work suggest that the present cavity approach based on the concept of cyclic heating and annealing is feasible.

\section{Conclusion and discussion}
\label{sec:discussion}

As a summary, in this paper we have calculated the thermodynamic properties
of a spin-glass model by combining the physical idea of
repeated heating-annealing and the cavity approach of M{\'{e}}zard and
Parisi \cite{Mezard-Parisi-2001}. We have assumed that, during a cyclic annealing
experiment, all the thermodynamic (or macroscopic) states of the spin-glass
system at a given low temperature $T$ will be reachable, but with different
frequencies which decrease exponentially with the free energy values of the
macroscopic states. By using this free energy Boltzmann distribution and by
using the Bethe-Peierls approximation, the grand free energy of the spin-glass
system can be calculated as a function of a reweighting parameter $y$; and from the knowledge of the grand free energy, the mean free energy, energy, and entropy of a macroscopic state of the system can also be obtained. For the $\pm J$ spin-glass model on a random regular graph of vertex degree $K=6$, the theoretical predictions of the present work are in good agreement with the results of earlier simulational and numerical calculations.

For the $\pm J$ model at $\beta=1.25$, we found that the complexity $\Sigma$ becomes negative before $y$ reaches $\beta$. Therefore, the thermodynamic properties of the system are contributed by those macroscopic states which have the global minimal free energy density. However, it may exist other systems for which the complexity is still positive even when the reweighting parameter $y$ reaches $\beta$. For such systems, the thermodynamic properties are not determined by those macroscopic states of the minimal free energy density, but by a set of $`$metastable' macroscopic states. On the one hand, these metastable macroscopic states have a higher free energy density; on the other hand, the number of such macroscopic states greatly exceeds the number of macroscopic states of the global minimal free energy density. In the competition between these two factors, the $`$metastable' macroscopic states may win. Then the system will reach one of these metastable macroscopic states almost surely in each round of the temperature annealing experiment. We will work on a model system with many-body interactions to check whether this is really the case.

As in the work of Ref.~\cite{Mezard-Parisi-2001}, the present theoretical treatment also assumes that the configurational space of the spin-glass system breaks into exponentially many macroscopic states, but there are no further organizations of these macroscopic states; and it is assumed that each macroscopic state is ergodic. As pointed out in Ref.~\cite{Montanari-etal-2004}, this first-step replica-symmetry-broken (1RSB) cavity solution might be unstable.
Two types of instability is conceivable. The type I instability concerns with
possible clustering of macroscopic states into $`$super-macroscopic' states;
the type II instability is  caused by splitting of each macroscopic state into
many $`$sub-macroscopic' states. To account for the further clustering of
macroscopic states appears to be relatively easy in the present mean-field framework:  for each cluster of macroscopic states, a vertex has a probability profile concerning its cavity magnetization; this probability profile is
different in different clusters, and we can introduce a probability distribution of probability profiles to characterize this variation [H.~Zhou, submitted to Comm.~Theor.~Phys. (Beijing)]. On the hand, to take into consideration the possibility of splitting of a macroscopic state, maybe one has to introduce other reweighting parameters in the mean-field theory. This is an important issue waiting for further explorations.

Another important issue is related to the updating of the probability profiles through the iterative equation (\ref{eq:rho}) or (\ref{eq:gfe-single-05}). In the present paper, we used Metropolis importance sampling technique to get an updated probability profile of cavity magnetization. This method appears to be rather precise but time-consuming. If faster algorithms with comparable precision could be constructed, it will be highly desirable.

\section*{Acknowledgement}

The numerical simulations reported in this paper were carried out at the
PC clusters of the State Key Laboratory for Scientific and Engineering 
Computing, CAS, China.

\section*{Appendix: On an explicit expression for the mean energy density of a spin-glass system}
\label{sec:appendix}

In the main text we did not give an explicit formula for the mean energy density $\epsilon$ of a spin-glass system. Instead, the mean energy density was calculated from first knowing the mean free energy density of the system [see, e.g., Eq.~(\ref{eq:mean-energy})]. In this appendix, we derive an explicit expression for the mean energy density $\epsilon$. The assumptions used in this derivation will be clearly pointed out.

\subsection*{A. Energy contribution of adding $K-1$ edges}
\label{subsec:appendix-a}

Consider again the two systems, ${\cal G}_K\bigl(N, 2(K-1) \bigr)$ in Fig.~\ref{fig:cavity01}A and ${\cal G}_K(N)$ in Fig.~\ref{fig:cavity01}B.
(For notational simplicity, let us hereafter referr to these two systems as system 'a' and 'b', respectively.) As was mentioned in Sec.~\ref{subsec:NtoN}, for a given configuration the energy difference between these two systems is $\Delta H_2$. When averaged over all macroscopic states, the mean total energy of the system 'a' is
\begin{equation}
\label{eq:Ea-01}
\bigl\langle \langle H^a \rangle_\alpha \bigr\rangle=
\frac{
 \sum\limits_{\alpha} \exp\bigl(-y F_\alpha^a\bigr)
\frac{
\sum_{\vec{\sigma}\in \alpha} H^a(\vec{\sigma}) \exp\bigl(-\beta H^a(\vec{\sigma}) \bigr)
}
{
\sum_{\vec{\sigma}\in \alpha} \exp\bigl(-\beta H^a(\vec{\sigma})\bigr)
}
}
{
\sum\limits_\alpha \exp\bigl(-y F_\alpha^a\bigr)
} \ ,
\end{equation}
where $F_\alpha^a$ is the total free energy of macroscopic state $\alpha$ of system 'a', and $H^a(\vec{\sigma})$ is the the configurational energy of system 'a'. Similarly, the mean total energy of the system 'b' is
\begin{equation}
\label{eq:Eb-01}
\bigl\langle \langle H^b \rangle_\alpha \bigr\rangle=
\frac{
 \sum\limits_{\alpha} \exp\bigl(-y F_\alpha^a - y \Delta F_2 \bigr)
\frac{
\sum_{\vec{\sigma}\in \alpha} \bigl[H^a(\vec{\sigma})+ \Delta H_2\bigr] \exp\bigl(-\beta H^a(\vec{\sigma}) - \beta \Delta H_2 \bigr)
}
{
\sum_{\vec{\sigma}\in \alpha} \exp\bigl(-\beta H^a(\vec{\sigma}) - \beta \Delta H_2 \bigr)
}
}
{
\sum\limits_\alpha \exp\bigl(-y F_\alpha^a - y \Delta F_2 \bigr)
} \ ,
\end{equation}
where $\Delta F_2$ is defined by Eq.~(\ref{eq:FreeEnergy01}). The difference of these two mean energy expressions is
\begin{equation}
\label{eq:V1V2}
\bigl\langle \langle H^b \rangle_\alpha \bigr\rangle -
\bigl\langle \langle H^a \rangle_\alpha \bigr\rangle
= V_1 + V_2 \ .
\end{equation}
In the above equation,
\begin{eqnarray}
V_1 &=& \frac{
\sum\limits_{\alpha} \exp\bigl(-y F_\alpha^a - y \Delta F_2 \bigr)
\frac{
\sum_{\vec{\sigma}\in \alpha}  \Delta H_2 \exp\bigl(-\beta H^a(\vec{\sigma}) - \beta \Delta H_2 \bigr)
}
{
\sum_{\vec{\sigma}\in \alpha} \exp\bigl(-\beta H^a(\vec{\sigma}) - \beta \Delta H_2 \bigr)
}
}
{
\sum\limits_\alpha \exp\bigl(-y F_\alpha^a - y \Delta F_2 \bigr)
} \ , \label{eq:V1m1} \\
&=& \sum\limits_{(i,j)\in e_1} \bigl\langle \Delta E^{(i j)} \bigr\rangle \ ,
\label{eq:V1}
\end{eqnarray}
where
\begin{equation}
 \label{eq:dEij}
\bigl\langle \Delta E^{(i j)} \bigr\rangle=-J_{i j}
\frac{
\int {\rm d} m_i \int {\rm d} m_j {\cal P}_i(m_i) {\cal P}_j(m_j) e^{-y \Delta F^{(i j)} } \frac{v_{i j} + m_i m_j}{1+v_{i j} m_i m_j}
}{
\int {\rm d} m_i \int {\rm d} m_j {\cal P}_i(m_i) {\cal P}_j(m_j) e^{-y \Delta F^{(i j)} }
} \ .
\end{equation}
In going from Eq.~(\ref{eq:V1m1}) to Eq.~(\ref{eq:V1}) we have used the Bethe-Peierls approximation both at the level of microscopic configurations [Eq.~(\ref{eq:factorization})] and at the level of macroscopic states [Eq.~(\ref{eq:factorization-magnetization})]. 

The term $V_2$ in Eq.~(\ref{eq:V1V2}) is equal to
\begin{equation}
 V_2=\frac{
 \sum\limits_{\alpha} \exp\bigl(-y F_\alpha^a - y \Delta F_2 \bigr)
\frac{
\sum_{\vec{\sigma}\in \alpha} H^a(\vec{\sigma})
\exp\bigl(-\beta H^a(\vec{\sigma}) - \beta \Delta H_2 \bigr)
}
{
\sum_{\vec{\sigma}\in \alpha} \exp\bigl(-\beta H^a(\vec{\sigma}) - \beta \Delta H_2 \bigr)
}
}
{
\sum\limits_\alpha \exp\bigl(-y F_\alpha^a - y \Delta F_2 \bigr)
}
- \frac{
 \sum\limits_{\alpha} \exp\bigl(-y F_\alpha^a\bigr)
\frac{
\sum_{\vec{\sigma}\in \alpha} H^a(\vec{\sigma}) \exp\bigl(-\beta H^a(\vec{\sigma})\bigr)
}
{
\sum_{\vec{\sigma}\in \alpha} \exp\bigl(-\beta H^a(\vec{\sigma})  \bigr)
}
}
{
\sum\limits_\alpha \exp\bigl(-y F_\alpha^a \bigr)
}
\end{equation}
At this point, let us introduce two assumptions:
\begin{enumerate}
\item[(i)] The distribution of the configurational energy of the cavity system 'a' and that of the energy increase  can be regarded as mutually independent within a macroscopic state.
\item[(ii)] The distributions of the free energy of the cavity system 'a' and that of the free energy increase  can also be regarded as mutually independent.
\end{enumerate}
If these two assumptions hold true, then we have
\begin{equation}
 \label{eq:V2zero}
V_2 =  0 \ .
\end{equation}

\subsection*{B. Energy contribution of adding two vertices and $2 K -1$ edges}
\label{subsec:appendix-b}

Now consider the two systems, ${\cal G}_K\bigl(N, 2(K-1) \bigr)$ in Fig.~\ref{fig:cavity01}A (system 'a') and ${\cal G}_K(N+2)$ in Fig.~\ref{fig:cavity01}C (system 'c'). Following the same analytical procedure as given in the preceding subsection, we find that, if the two assumptions (i) and (ii) listed above are valid when going from system 'a' to system 'c', then the mean total energy of system 'c' is related to that of system 'a' by the following simple equation
\begin{equation}
\label{eq:V1V2b}
\bigl\langle \langle H^c \rangle_\alpha \bigr\rangle -
\bigl\langle \langle H^a \rangle_\alpha \bigr\rangle
= \bigl\langle \Delta E^{(i_0 j_0)} \bigr\rangle
+ \sum\limits_{j_s\in \partial i_0 \backslash j_0} \bigl\langle \Delta E^{(i_0 j_s)} \bigr\rangle + \sum\limits_{i_s\in \partial j_0 \backslash i_0} \bigl\langle \Delta E^{(j_0 i_s)} \bigr\rangle \ .
\end{equation}
(Notice that in calculating $\bigl\langle \Delta E^{(i_0 j_s)} \bigr\rangle$ through Eq.~(\ref{eq:dEij}), $m_{i_0}$ should be the cavity magnetization of vertex $i_0$ in the absence of vertex $j_s$.).

Combining Eq.~(\ref{eq:V1V2}) and Eq.~(\ref{eq:V1V2b}), we finally obtain that the mean free energy density of the system is
\begin{equation}
 \label{eq:epsilon-final}
\epsilon(\beta;y)=\frac{K}{2} \overline{\bigl\langle \Delta E^{(i j)} \bigr\rangle } \ .
\end{equation}

The same analytical procedure can be applied to calculate the mean free energy density. Under the assumption (ii) of the preceding subsection, we obtain that the mean free energy density $f$ is equal to
\begin{equation}
 \label{eq:f-another-final}
f(\beta;y)=\overline{\bigl\langle \Delta F^{(i)} \bigr\rangle}
-\frac{K}{2}\overline{\bigl\langle \Delta F^{(i j)} \bigr\rangle} \ ,
\end{equation}
which is just the same as Eq.~(\ref{eq:free-energy-1RSB}).

Based on the assumption of exponentially many macroscopic states, it was argued in Ref.~\cite{Mezard-Parisi-2001} that, the assumption (ii) of the preceding subsection should be valid under the Bethe-Peierls approximation. In other words, the joint distribution of the free energy of the old cavity system $F$ and the free energy increase $\Delta F$ is factorized. The consistence between Eq.~(\ref{eq:f-another-final}) and Eq.~(\ref{eq:free-energy-1RSB}) also confirmed this point. Since each macroscopic state ifself contains an exponential number of microscopic configurations, we believe that the assumption (i) is also valid in the Bethe-Peierls mean-field framework. The mean energy density of the $\pm J$ spin-glass model as shown in Fig.~\ref{fig:result01} was calculated using Eq.~(\ref{eq:epsilon-final}), and the output was in good agreement with earlier simulation result. Maybe one can further check this point by working on some simple models which are exactly solvable both analytically and computationally. The maximum matching problem as studied in Refs.~\cite{Zhou-Ouyang-2003,Zdeborova-Mezard-2006} could be a good candidate model system.


\end{document}